\documentclass[english]{article}
\usepackage[T1]{fontenc}
\usepackage[latin9]{inputenc}
\usepackage{amsmath}
\usepackage{amssymb}

\makeatletter

\newcommand{\noun}[1]{\textsc{#1}}

\newcommand{\lyxaddress}[1]{
	\par {\raggedright #1
	\vspace{1.4em}
	\noindent\par}
}

\usepackage{babel}

\usepackage{babel}

\makeatother

\usepackage{babel}
\begin{document}
\title{How fast is a quantum jump?}
\author{L. de la Peña, A. M. Cetto\thanks{Corresponding author. Email: ana@fisica.unam.mx}\ \ and
A. Valdés-Hernández \\
}
\maketitle

\lyxaddress{Instituto de Física, Universidad Nacional Autónoma de México, 04510
Mexico}
\begin{abstract}
A proposal is put forward for an estimate of the duration of a transition
between atomic states. The proposal rests on the consideration that
a resonance of the atomic electron with modes of the zero-point radiation
field of Compton's frequency is at the core of the phenomenon. The
theoretical result, given essentially by the expression $(\alpha\omega_{C})^{-1}$,
where $\alpha$ is the fine structure constant and $\omega_{C}$ the
Compton angular frequency for the electron, lies well within the range
of the recently experimentally estimated values of the order of attoseconds
(10$^{-18}$ s).

\end{abstract}

\section{Introduction}

It is in a way puzzling that the physics involved in atomic quantum
jumps (or single atomic transitions) has remained in almost complete
darkness, the more so considering the crucial role spectroscopy has
played for more than a century, and the impressive advances in both
theoretical and experimental quantum physics. Attention to this intriguing
subject has presumably been hindered for a long time by the masterful
dogma of the instantaneous character of atomic transitions postulated
by Bohr \cite{Bohr}---and bitterly opposed by Schrödinger \cite{SchrQJ1,SchrQJ2}.
One can still come across articles negating quantum jumps---and any
other kind of discontinuities, for that matter (e.g. \cite{Jadczyk})---or
taking them as a sudden increase of our knowledge of the system (e.
g., \cite{Zeh,Plenio}) rather than a physical phenomenon.

For atomic and molecular spectroscopists it is clear that quantum
jumps exist; this is part of their daily bread. Most spectroscopists
are also aware that the time involved in a transition is finite but
very short; so short indeed that the Franck-Condon principle applies,
which sets an upper limit to their duration, on the order of femtoseconds
($10^{-15}$ s). Because atomic transitions are so fast, up to recently
they were considered ``instantaneous'', this term being taken by
some in the rigorous sense, and by others as meaning ``in an unmeasurably
(or unobservably) short time''. This picture, however, is changing
thanks to recent calculational and experimental work, notably using
attosecond spectroscopy applied to photoionization \cite{Paz,Oss}.
Photoionization experiments in bulk materials are known to involve
electronic correlations, which makes it difficult to ascertain the
time it takes for one atom to lose one electron. With this caveat,
chronoscope measurement of the times involved in the photoelectric
effect assigns to the primary photoexcitation process a duration on
the order of $10^{-17}$ s \cite{Oss2}. Further, although not directly
comparable to (natural) atomic transitions, recent experimental work
with an artificial atom (a superconducting circuit consisting of two
hybridized qubits on a chip) in which a quantum jump is intercepted
and reverted by means of an electric pulse, seems to confirm Schrödinger's
intuition that the evolution of the jump itself is continuous and
needs a finite time to take place \cite{Minev}. This is in line with
Schulman's definition of ``jump time'' as the time scale such that
perturbations occurring at intervals of this duration affect the transition
\cite{Schul}. Based on his definition, Schulman's own estimate made
in terms of the ``Zeno time'' (related to the second moment of the
Hamiltonian) and the natural lifetime, results however in a much shorter
time than the experimental estimates, as short as $10^{-20}$ s for
atomic transitions. 

The various computational and experimental estimates have contributed
to establish the existence of (finite-time) quantum jumps, and have
apparently set tighter bounds on their duration. The basic physics
behind the process, however, has not been clarified, so the question
remains: what is it that determines the duration of a transition? 

In the present work we attempt to throw light on this question via
a theoretical analysis that does not rely on specific experimental
settings. We do so by invoking the existence of the electromagnetic
zero-point radiation field \textsc{(zpf}) and applying the conventional
approach followed in stochastic electrodynamics (\textsc{sed}) to
the specific problem of the dynamics of the electron during a transition.
We start by recalling Schrödinger's work on the zitterbewegung as
a rapid oscillation of the Dirac electron, and appeal to \textsc{sed}
to identify it as a result of its resonance with the Compton-frequency
components of the \textsc{zpf}. The stationary solution of the equation
of motion for the rapidly oscillating electron corresponds to the
zitterbewegung; the transient solution, in its turn, describes the
dynamics of the transition between states. The decay time associated
with the transient solution, which we propose to take as an approximate
measure of the transition time, is expressed in terms of universal
constants and its value is of the order of 10$^{-18}$ s. 

\section{The guiding premise \label{prem}}

To put the discussion on track we start by recalling the source of
the zitterbewegung as disclosed by Schrödinger \cite{SchZit} in his
revision of the properties of the free particle in Dirac's theory
of the electron. This will signal the importance of the Compton-frequency
modes of the \textsc{zpf} for the dynamics of the electron, and pave
the way for their consideration as a central element in the transition.

A well-known result in Dirac's theory of the free electron is that
the velocity operator is $c\hat{\alpha}$, where $\hat{\alpha}$ is
one of the matrices in the Dirac theory and $c$ stands for the velocity
of light in vacuum; this is expressed in the equation $\dot{x}=c\hat{\alpha}$.
As shown by Schrödinger \cite{SchZit}, the Heisenberg equation of
motion gives for $\dot{x}$ in terms of the canonical momentum $p$
acquired by the particle with energy $E$ at time $t,$ the expression
(for simplicity we use one-dimensional notation)
\begin{equation}
\frac{E}{c^{2}}\dot{x}=p-\left(p-\frac{E}{c^{2}}\dot{x}(0)\right)e^{-i2Et/\hbar}.\label{10hp}
\end{equation}
To arrive at this result, Schrödinger considered both the energy and
the momentum of the free particle as having a constant value; the
coefficient $E/c^{2}$ stands for the mass of the particle.

Let us now compare this result with the conventional definition of
the canonical momentum $p$ for a particle in the presence of an electromagnetic
field $A$, namely 
\begin{equation}
m\dot{x}=p-\frac{e}{c}A.\label{12hp}
\end{equation}
The comparison may seem unwarranted, since from the viewpoint usually
adopted to read the quantum-mechanical formalism, no electromagnetic
field exists other than any expressly recognized external field, whence
for the free particle one should take $A=0$. Here, by contrast, we
propose to explore the possibility of taking the comparison at face
value. This means that the quantity $\left(p-\frac{E}{c^{2}}\dot{x}(0)\right)e^{-i2Et/\hbar}$
in Eq. (\ref{10hp}) represents the effect of an acting electromagnetic
field of very high frequency, essentially twice Compton's frequency
$\omega_{C}=mc^{2}/\hbar$, and an amplitude of the order of Compton's
wavelength $\lambda_{C}=\hbar/mc$, as follows from an integration
of Eq. (\ref{10hp}). This extra oscillation of the (free) particle
predicted by the Dirac equation exhibits the zitterbewegung as a real
helicoidal motion with velocity $c$ around the particle trajectory
\cite{SchZit}. Incidentally, notice that Schrödinger had real particle
trajectories in mind.

The above identification acquires full sense within the framework
of \noun{sed,} which is based on the premise that the electron is
permanently embedded in the \noun{zpf }(for different reviews of \noun{sed}
see, e. g., \cite{Boy,Dice,TEQ,Boyer19}). The (nonrelativistic) \noun{sed}
equation of motion for a particle of charge $e$ and mass $m$, subject
to an external force $\boldsymbol{f}(\boldsymbol{x})$, is the corresponding
Abraham-Lorentz equation of classical electrodynamics, extended to
include the \textsc{zpf.} It reads (here the dynamical variables are
c-numbers) 
\begin{equation}
m\ddot{\boldsymbol{x}}=\boldsymbol{f}(\boldsymbol{x})+m\tau\dddot{\boldsymbol{x}}+e\left[\boldsymbol{E}(\boldsymbol{x},t)+\frac{\dot{\boldsymbol{x}}}{c}\times\boldsymbol{B}(\boldsymbol{x},t)\right].\label{16hp}
\end{equation}
The term $m\tau\dddot{\boldsymbol{x}}$ stands for the (nonrelativistic)
expression for the force due to radiation damping, with $\tau=2e^{2}/3mc^{3}\sim$
10$^{-23}$ s for the electron. The term within brackets is the Lorentz
force due to the radiation field. In consonance with the nonrelativistic
treatment, the field is normally taken in the dipole approximation,
whence (we resume one-dimensional notation, for simplicity), 
\begin{equation}
m\ddot{x}=f(x)+m\tau\dddot{x}+eE_{x}(t).\label{17hp}
\end{equation}
For the treatment of the majority of atomic problems in quantum mechanics
this approximation has proven legitimate and sufficient; even the
radiative lifetimes and (nonrelativistic) radiative corrections are
correctly obtained under this approximation \cite{Dice,TEQ}.

Using canonical variables, for which $\dot{p}=f(x)$, and writing
in the Coulomb gauge $E_{x}=-\frac{1}{c}\partial A_{x}/\partial t$,
integration of Eq. (\ref{17hp}) leads to 
\begin{equation}
m\dot{x}=p+m\tau\ddot{x}-\frac{e}{c}A_{x}(t)=p-\frac{e}{c}A_{T}(t).\label{18hp}
\end{equation}
In the second equality, $A_{T}(t)$ represents the \emph{total} radiation
field in the $x$ direction, $A_{T}=A_{x}-(2e/3c^{2})\ddot{x}$. In
the absence of external radiation fields, this reduces to the \textsc{zpf}
plus particle radiation (a more detailed discussion can be seen in
chapter 6 of Ref. \cite{Dice}).

Now the comparison of Eqs. (\ref{10hp}) and (\ref{18hp}) is immediate,
(\ref{18hp}) being the nonrelativistic (\textsc{sed}) counterpart
of (\ref{10hp}). This comparison suggests that---as is frequently
the case---the relativistic treatment of the quantum problem automatically
includes (some of) the effects of the \textsc{zpf} on the motion of
the particle, even if this field is not expressly introduced. In other
words, the quantum description already contains information about
the presence of the \noun{zpf}.

According to this discussion, within the framework of \noun{sed} the
oscillations manifested as zitterbewegung are the result of a resonant
interaction of the particle with the components of the \textsc{zpf}
having a frequency of the order of Compton's frequency.\footnote{The relativistic frequency is $2mc^{2}/\hbar$ due to the simultaneous
consideration of both the positive and negative energies. In the nonrelativistic
case the Compton frequency $mc^{2}/\hbar$ is a more natural limit
for the descriptive capacity of the theory. More detailed discussions
about the electron resonance at Compton's frequency can be seen in
Refs. \cite{NosSpin,Hest2}.} This suggests a prominent role for $\omega_{C}$ in the dynamical
behaviour of the atomic electron, and gives a clue for understanding
other dynamical effects, even in the nonrelativistic scenario, as
will be shown in the following section in relation with atomic transitions.

\section{How fast is a quantum jump?}

We turn now to our task of estimating an order of magnitude for the
time it takes the atomic electron to make a transition between states,
guided by the above considerations. The gist of our argument is, as
stated above, the acknowledgement that the electron resonates with
the modes of the \textsc{zpf} of Compton's frequency, in addition
to the (slow) motion impressed upon it by the external forces and
the low frequency components of the \noun{zpf}. We shall take the
simplest nonrelativistic approach to tackle the problem. Therefore,
we apply Eq. (\ref{17hp}) to the actual position variable---which
we denote now as $x'(t)$ instead of $x(t)$, with $x'=x+z$---and
separate the terms corresponding to the slow motion, represented by
$x(t)$, from those associated with the (normally) small but rapid
motion, represented by $z(t)$.

A Taylor series expansion up to first order in $z$, of $x'(t)$ around
$x(t)$, gives for the equation describing the slow motion 
\begin{equation}
m\ddot{x}=f(x)+m\tau\dddot{x}+eE'(t),\label{10qj}
\end{equation}
where $E'(t)$ represents the \noun{zpf} except for its high-frequency
modes. This is equivalent to the usual \textsc{sed} equation of motion
in the long-wavelength approximation, and, as said above, it serves
in general to solve nonrelativistic atomic problems. In particular,
the correct radiative lifetimes are obtained both for spontaneous
and induced transitions, as a result of the resonant response of the
atomic electron to the (long-wavelength) modes of the radiation field,
the \noun{zpf} included \cite{TEQ,AB}.

To study the dynamics of the transition itself we need the equation
of motion for the fast variable $z(t)$, which is obtained by collecting
the remaining terms not contained in Eq. (\ref{10qj}) and including
a force term $-m\omega_{C}^{2}z$ to account for the resonance of
the electron at the Compton frequency,

\begin{equation}
m\ddot{z}=zf'(x)-m\omega_{C}^{2}z+eE_{C}(t)+m\tau\dddot{z},\label{12qj}
\end{equation}
where $E_{C}(t)$ stands for the high-frequency modes of the \noun{zpf}.
The term $zf'(x)$ is small compared with the remaining force terms
and can be neglected; we are thus left with 
\begin{equation}
m\ddot{z}=-m\omega_{C}^{2}z+m\tau\dddot{z}+eE_{C}(t).\label{13qj}
\end{equation}

The stationary (forced) solution of this inhomogenous equation represents
the zitterbewegung, which takes place during the entire life of the
electron, thanks to the permanent action of the high-frequency \noun{zpf}
modes represented by $E_{C}$, as discussed in Sect. \ref{prem}.
In addition, the homogeneous part of the equation admits a transient
solution $z_{\mathrm{tr}}(t)$. Writing to first order in $\tau$
\begin{equation}
\omega_{C}\sqrt{1+i\tau\omega_{C}}\simeq\omega_{C}+\frac{1}{2}i\tau\omega_{C}^{2},\label{30qj}
\end{equation}
we obtain 
\begin{equation}
z_{\mathrm{tr}}(t)=z_{0}\exp({i\omega_{C}\sqrt{1+i\tau\omega_{C}}t})+\text{c.c.}\simeq e^{-\tau\omega_{C}^{2}t/2}\left(z_{0}e^{i\omega_{C}t}+z_{0}^{*}e^{-i\omega_{C}t}\right),\label{28qj}
\end{equation}
where the constants of integration $z_{0}$ and $z_{0}^{*}$ are to
be determined by the initial value $z_{\mathrm{tr}}(0)$.

We propose to identify the irreversible change in the state of motion
described by Eq. (\ref{28qj}) with a transition between (atomic)
states, $z_{\mathrm{tr}}(0)$ giving an idea of the distance traveled
during the transition. The change of state implies a well-defined
increase or decrease in the energy of the system, normally accompanied
by an absorption or emission of radiation. It does not, however, entail
in principle any discontinuity in the trajectory---as the image implicit
in the notion of a ``quantum jump'' seems to suggest.

The characteristic time $T_{\textrm{tr}}$ for the decay, which according
to this proposal can be taken as a measure of the duration of the
transition between states, is (using $\tau=2e^{2}/3mc^{3}$ and $\omega_{C}=mc^{2}/\hbar$)
\begin{equation}
T_{\textrm{tr}}\simeq\frac{2}{\tau\omega_{C}^{2}}=\frac{3\hbar^{2}}{e^{2}mc}.\label{32qj}
\end{equation}

It is remarkable that $T_{\textrm{tr}}$ becomes expressed in terms
of the four fundamental constants $e,m,c$, $\hbar$, meaning that
its order of magnitude can be evaluated from simple dimensional considerations.
Noting that $\tau\omega_{C}=2\alpha/3,$ where $\alpha=e^{2}/\hbar c$
is the fine-structure constant, we get for $T_{\textrm{tr}}$ the
alternative expressions 
\begin{equation}
T_{\textrm{tr}}=\frac{3}{\alpha\omega_{C}}=\frac{3T_{C}}{2\pi\alpha},\label{34qj}
\end{equation}
which shows that the transition time is larger by two orders of magnitude
than the Compton time $T_{C}.$ With $\lambda_{C}$ = 2.43$\times10^{-10}$
cm, we have $T_{C}=8.1\times10^{-21}$ s, which gives 
\begin{equation}
T_{\textrm{tr}}\simeq65.4\,T_{C}=0.53\times10^{-18}\text{ s.}\label{36qj}
\end{equation}
The theoretical value thus obtained for the jumping (transition) time
lies well within the range of recent empirical evaluations. A more
elaborate, relativistic treatment would most certainly produce more
precise results. 

Finally, it is interesting to note that the times corresponding to
the (inverse of the) frequencies of atomic spectral lines lie approximately
between $0.3\times10^{-18}$ s and $0.3\times10^{-17}$ s, which would
indicate that the (emitted or absorbed) radiation field performs about
one oscillation during the transition process.

Acknowledgments: The authors acknowledge financial support from DGAPA-UNAM
through project PAPIIT IN113720.

\end{document}